\documentclass[]{elsart}
\usepackage{bm}
\usepackage{epsfig}
\begin{document}
\begin{frontmatter}

\title{Relativistic nature of the EMC-effect}

\author[Vladivostok]{Alexander Molochkov\thanksref{AVH}}
\address[Vladivostok]{Far Eastern National University, 690950 Vladivostok, Russia}
\thanks[AVH]{Partially supported by the Alexander von Humboldt Foundation, Germany}

\begin{abstract}
The deep inelastic scattering of leptons off nuclei is studied within the
the Bethe-Salpeter formalism. It is shown that nuclear short-range structure
can be expressed in terms of the nucleon
structure functions and four-dimensional Fermi motion of the nucleons.
The four-dimensional Fermi motion broadens the bound nucleon localization area,
what leads to the
observation of the nucleon structure change in nuclei -- EMC effect.
The $\rm ^4He$ to deuteron
structure functions ratio is found in good agreement with experimental data.
It is shown that the pattern of the ratio is defined by dynamical properties
of the nucleon structure and four-dimensional geometry of
the bound state.
\end{abstract}
\end{frontmatter}

The European Muon Collaboration (EMC) demonstrated in the experiments on
deep inelastic scattering (DIS) of leptons off nuclei that
nuclear environment modifies short range nucleon structure~\cite{aub83}.
The modifications were observed as deviations from unity of the nuclear and deuteron
DIS cross-sections ratio $R^A$= $2\sigma^A$/$(A\sigma^{\rm D})$
to the values smaller than one.
Since it was widely accepted that soft momentum nucleon-nucleon interaction in nuclei
cannot modify a hard momenta distribution of nucleon partons,
this phenomena was very unexpected.
Later this effect was studied in a wide kinematic range in
many experiments~(c.f. review \cite{arneodo}).
A wide variety of models was proposed
to explain these modifications (c.f. reviews~\cite{arneodo,reviews}).
In different kinematic ranges the oscillations were
considered as the different effects that are
shadowing, antishadowing, EMC effect, and Fermi motion.
No one of the models provided a quantitative explanation of the effect
in the whole kinematic range, what led to
the conclusion of the review~\cite{arneodo} that origin of
the EMC-effect stays unclear.

So, the EMC-effect remains a topical subject up to now.
Experimental study of nuclear hard structure provided opportunities to find
important regularities that can give
additional constrains on the models of the
EMC-effect.
Study of the
EMC-effect dependence
on the atomic number of the nucleus (A-dependence)
performed in the SLAC experiments~\cite{gomez}
showed that amplitude of the effect does not saturate with increase of A.
The consequent analysis of the world data for the $2\sigma^A/(A\sigma^{\rm D})$ ratio
made in the paper~\cite{sm95} uncovered universality in the EMC-effect
A-dependence for the all kinematic ranges.
An another important result was obtained in the hard $pA$ scattering performed in
FNAL~\cite{DY}.
Analysis of the nuclear and deuteron Drell-Yan cross-sections ratio
showed no excess of the antiquark component in nuclei.

These results provided a
basis for the critics of the most obvious explanations of the EMC-effect
that were proposed by the binding~\cite{xresc}
and mesonic exchange models~\cite{MEC}.
In the paper~\cite{miller} it was stressed that
due to the Hugenholtz-van Hove
theorem~\cite{hugenholtz} the nuclear binding effect in DIS is defined by the
mass defect in the nucleus. Thus, the binding effect has to
saturate with increasing of $A$, while the systematic experimental study
of the EMC-effect $A$-dependence~\cite{gomez} does
not show such saturation.
 Results of the Drell-Yan experiment~\cite{DY} prove
that the mean mesonic field in the nucleus,
which presumably consist of $q \bar q$ fields, also cannot
explain the EMC-effect~\cite{miller}.

This critics led to the conclusion that nucleon energy-momentum
change due to the binding effects cannot be responsible alone for
the EMC-effect. The analysis performed within the QMC
model~\cite{thomas03} pointed that the EMC-effect cannot be
explained without introducing the hypothesis that nucleon
structure is changed in nuclear media. This conclusion is
consistent with the previously obtained resume of the calculation
based on the quasi-potential approach~\cite{liuti} and supported
recently by the light front analysis~\cite{miller03}.

However, despite this critics, the binding model provided an important signal
about nucleon structure change in nuclei.
This model assumes the bound nucleon mass shift $m\to m^*=m-\epsilon$,
where $\epsilon$ is the binding energy of the nucleon.
Due to the uncertainty relation the mass-shift changes the observed radius of
the nucleon localization area in the four-dimensional ($4$D) space~\cite{mynucl}.
Thus, this mass-scale shift leads to the change of the quark confinement 4D-radius
and, hence, to distortion of the partonic distribution inside the nucleon.
This explanation coincides with the $Q^2$-rescaling model~\cite{Q2resc},
which explains the EMC-effect as a change of the quark confinement radius
in the bound nucleon.
In that way, the explanations of the EMC--effect that are proposed by the
binding and $Q^2$-rescaling models can be reduced to the
distortions of the
nucleon $4$D-structure in the nuclei.  Thus, a fully covariant $4$D-treatment
is essential for understanding of this phenomenon.

In the present letter I would like to focus on the bound state
relativistic properties that are important for understanding of the EMC-effect.
To clarify the role played by the
relativistic effects it is essential to consider the bound state within
an explicitly covariant approach that can be developed
in the relativistic field theory framework.
Within the covariant field theory the space-time distribution of the nucleons inside the
nucleus is defined by the following amplitude
\begin{equation}
\phi(x_1,\dots x_n)=\langle 0|T\psi(x_1)\dots \psi(x_n)|A \rangle,
\label{BSvertex}\end{equation}
where $\psi$ denotes the nucleon field operators, the four vectors $x_i$ define
positions of the bound nucleons in the space-time.
Hence, nucleons
inside the nucleus are separated
not only by the three dimensional space-intervals ${\bf r}_{i,j}={\bf x}_j-{\bf x}_i$,
but by the time-intervals ${\tau}_{i,j}={x_0}_j-{x_0}_i$ as well.
The separated in time constituents is
the specific relativistic bound state property
(it is called $\tau$-shift bellow) that
was criticized from the very beginning of the relativistic
 bound state theory development.
The critical comments claimed that the shifted in time
constituents mean causality violation~\cite{tcritics}. In that way
the $\tau$-shift was considered as a non-physical property, which
is not reflected in observables and can be fixed to any value. It
stimulated development of the quasi-potential approaches where the
$\tau$-shift was fixed according to different external conditions,
which have to lead to equivalent results in calculations of
observable quantities (c.f. reviews~\cite{quasipot}). However,
this equivalence was proven to be false in the analysis of the
relativistic covariance of the quasi-potential approaches
performed in the paper~\cite{pascalutsa}.

To clarify the role played by the $\tau$--shift it is important to note that the
quoted above critics has sense only within classical limit, where
space-time positions of particles can be detected exactly.
A position of a quantum particle
in the space-time can be detected within the boundaries that are
defined by the uncertainties in detection of the energy and momentum
($\Delta E$ and $\Delta p$)
of the particle: $\Delta x\simeq 1/\Delta p, \,\,
\Delta t\simeq 1/\Delta E$.
Since the uncertainty in energy detection of the nucleon partons defined
by the total nucleon energy, $\Delta E_q=E_{\rm N}$; the boundaries
of the nucleon localization in time are defined as
$\Delta t_{\rm N}=1/E_{\rm N}$.
Since the bound nucleon is distributed in the space-time
inside the nucleus and energy of the nucleon is shifted from the mass-shell,
 the boundaries of the bound nucleon are extra blurred:
$\delta \Delta t \simeq 1/(E_{\rm N}-\Delta E_{\rm N}) - 1/E_{\rm N}$,
where $\Delta E_{\rm N}$ is the energy shift of the
nucleon due to the binding and Fermi motion. Thus, the causality violation
is unobservable if the $\tau$-shift is not larger than the extra blurring area
$\delta \Delta t$ defined by the energy-shift of the bound nucleon:
\begin{equation}
\tau \le \frac{\Delta E_{\rm N}}{E_{\rm N}(E_{\rm N}-\Delta E_{\rm N})}.
\label{causality}\end{equation}
If this expression holds true then the effects that come from the $\tau$--shift
cannot be considered as an evidence
of the causality violation.
In this case the complete relativistic picture of the nuclear effects
has to incorporate the
$\tau$-shift, which is lost in the conventional nuclear binding models
based on the quasi-potential approaches.

For calculation of the $\tau$--shift effects
an explicitly covariant field theory formalism is essential.
In the present paper
I use the approach that was
developed in the series of the publications~\cite{mynucl,approach}
on the base of the Bethe-Salpeter equation~\cite{BS}.

Let us consider the main line of the calculations.
Due to the optical theorem the amplitude
of the high-energy inclusive scattering is proportional to the imaginary part
of the amplitude for the
forward scattering $\langle A|T(J_{\mu} J_{\nu})|A \rangle$.
Within the Bethe-Salpeter formalism this matrix element is defined
by the space-time distributions~(\ref{BSvertex}) and
vacuum average of the T-product of the nucleon fields and
nucleon em-current~( see Ref. \cite{approach}).
Following this method we get the expression for the nuclear
DIS amplitude $W^{A}_{\mu\nu}$. Within the Bjorken limit ($Q^2\to \infty$)
all terms, except the relativistic impulse approximation term, will come to zero at
least as $1/(Q^2)^2$~\cite{approach}. The relativistic impulse approximation
term, where the lepton scatters off the single nucleon in the nucleus, has the
 following form:
\begin{equation}
W^{A}_{\mu\nu}(P_A,q)= \int\frac{d^4p}{(2\pi)^4}\frac{W^{\rm
N}_{\mu\nu}\left(p,q\right) f^{{\rm
N}/A}(P,p)}{\left(p^2-m^2\right)^2
(\left(P-p\right)^2-M_{A-1}^2)}. \label{fermi4d}\end{equation}
This expression gives the nuclear DIS amplitude in terms of the
off-mass-shell nucleon DIS amplitude $W^{\rm N}_{\mu\nu}$ and the
nucleon distribution function $f^{\rm N}(P_A,p)$. This
distribution function is defined by the
 amplitudes~(\ref{BSvertex}), and together with the denominator it composes
 the four-dimensional momentum distribution of the struck nucleon
 inside the nucleus carrying the total momentum
 $P_A=(M_A,{\bf P})$.
In that way Eq.~(\ref{fermi4d}) expresses the nucleon blurring that results
from the four-dimensional distribution of the nucleon inside the nucleus.
By analogy with the non-relativistic $3$D momentum distribution I will call it
four-dimensional Fermi motion.

Due to the four-dimensional integration in Eq.~(\ref{fermi4d}) actual
calculations require information about nucleon amplitude $W^{\rm N}_{\mu\nu}$
in the kinematic region of the off-mass-shell values of the
nucleon energy $p_0$ ($p_0^2 \neq {\bf p}^2 + m^2$).
The off-mass-shell behavior of $W^{\rm N}_{\mu\nu}$ is unobservable, since then
explicit microscopic calculations of the amplitude have no experimental
reference and strongly model dependent.
Thus, Eq.~(\ref{fermi4d}) has to be rewritten in terms of measurable
quantities such as the DIS structure functions of the physical nucleon, which
defined by the total DIS cross-section in the Bjorken limit
$\sigma^{\rm N}\propto F_2^{\rm N}$.
The most simple solution of this problem is provided by
the integration in Eq.~(\ref{fermi4d})
with respect to $p_0$.
Analytical properties of the integrand in Eq.~(\ref{fermi4d}) give a
way to do it explicitly.
Within the assumption that $W^{N}_{\mu\nu}(p,q)$ and $f^{N}(P_A,p)$ are regular
with respect to $p_0$ \footnote{Since $W^{N}_{\mu\nu}=Im_{q_0} T^{N}_{\mu\nu}$,
the first assumption is obvious.
The second assumption was proven in the papers~\cite{mynucl,phlprog}.}
the integrand  contains a second order pole corresponding to the struck
nucleon and a first oder pole corresponding to the spectator.
These poles lie in the different
half-planes of the complex plane $p_0$. So, we can choose one of the
singularities to perform the integration. To express the
nuclear hadron tensor in terms of the nucleon structure functions
it is necessary to choose the second
order pole ($p_0-E_{\rm N}$) that corresponds to the struck nucleon.
The result of the contour integration in vicinity of the second order pole in
Eq.(\ref{fermi4d}) is defined by the derivative of the pole residue
with respect to $p_0$ at the point $p_0=E_{\rm N}$.
Doing the integration and using the relation $W_{\mu\nu}(P,q)=F_2(x,Q^2)/x$
we get the following expression for the nuclear structure function $F_2^A$:
\begin{equation}
F_2^{A}(x)=
\int\frac{d^3p}{(2\pi)^3}
\left[\frac{M_A-E_{A-1}-{p_3}}{E_{\rm
N}}F_2^{\rm N}(x_{\rm N})- \frac{\Delta^{\rm N}_{A}}{E_{\rm N}}
x_{\rm N}\frac{d F_2^{\rm N}(x_{\rm N})} {dx_{\rm N}} \right]
\frac{f^{N/A}(M_A,{\bf p})}{8M_{A}E_{\rm N}E_{A-1}{\Delta^{\rm N}_{A}}^2},\label{F2A}
\end{equation}
where $x=-q^2/(2mq_0)$ is the nuclear Bjorken $x$ normalized on the nucleon mass,
$x_{\rm N}=xm/(E_{\rm N}-p_3)$ is the Bjorken $x$ of the struck nucleon,
$E_{\rm N}=\sqrt{m^2+{\bf p}^2}$ is the struck nucleon on-shell energy,
 $E^2_{A-1}={M^2_{A-1}+{\bf p}^2}$ is the on-shell energy of the nuclear residue,
 $\Delta^{\rm N}_{A}=M_A-E_{\rm N}-E_{A-1}$.

 The function $f^{N/A}(M_A,{\bf p})$ together with the denominator composes the
 three dimensional momentum
distribution of the struck nucleon inside the nucleus.
According to the normalization condition
for the Bethe-Salpeter vertex function this distribution satisfies the baryon
 and momentum sum rules~\cite{mynucl} and coincides with the usual nuclear
momentum distribution. Thus, the first term in Eq.(\ref{F2A}), which results from the
derivative of the propagator of the nuclear spectator,
expresses contribution of the conventional $3$D-Fermi motion.

The second term with $dF_2^{\rm N}(x_{\rm N})/dx_{\rm N}$ results from the
nucleon DIS amplitude derivative:
$$\frac{dW^{\rm N}_{\mu\nu}(p,q)}{dp_0}=
\frac{dx_{\rm N}}{dp_0}\left(\frac{1}{x_{\rm N}}\frac{dF^{\rm N}_2(x_{\rm N},Q^2)}
{dx_{\rm N}}-
\frac{F^{\rm N}_2(x_{\rm N},Q^2)}{x_{\rm N}^2}\right).$$
This expression shows the $\tau$--shift influence on the
observable quantities in nuclear DIS.
Its contribution is proportional to the coefficient $\Delta^{\rm N}_A$;
which, therefore,
characterizes contribution of the $\tau$-shift.
Due to the energy conservation $\tilde E_{\rm N}+E_{A-1}=M_A$, where
$\tilde E_{\rm N}$
is the off-mass-shell energy of the bound nucleon. Therefore, $\Delta^{\rm N}_A$
is equivalent to the to the total energy shift of the struck nucleon
due to the binding and Fermi motion $\Delta^{\rm N}_A=\tilde E_{\rm N}-E_{\rm N}$.
Thus, the $\tau$--shift satisfies the causality condition~(\ref{causality}).

It is important to note that the struck nucleon pole gives the
factor $1/{\Delta^{\rm N}_{A}}^2$, since then the contribution
from all other singularities (for example nucleon self-energy cut
or anti-nucleon pole) are suppressed at least as $(\Delta^{\rm
N}_{A}/M_A)^2$~\cite{approach}. Since the mean value of the energy
shift is small for all nuclei ($\Delta^{\rm N}_A/M_A \propto
O(10^{-2})$), Eq.(\ref{F2A}) provide nuclear structure function
with accuracy up to terms of order $(\Delta^{\rm
N}_{A}/M_A)^2\propto O(10^{-4})$.

\begin{figure}
\epsfxsize=70mm \epsfbox{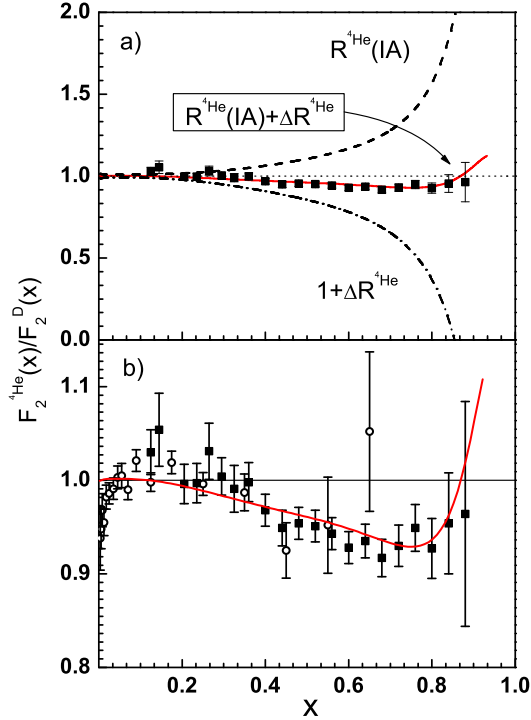}
\caption{\label{fig1}Ratio of the $^4$He and D structure functions.
a) Contributions to the ratio of the impulse approximation (dashed curve $R^{^4He}(IA)$)
 and
nucleon structure function derivative (dashed-dot curve $\Delta R^{^4He}$).
The full calculation (full curve) is the sum of these contributions. The experimental
values are shown by the dark squares~\protect\cite{gomez} and
 the light circles~\protect\cite{ama95}.}
\end{figure}

Numerical calculations of the contributions of the different terms
in Eq.(\ref{F2A}) to the ratio $R^{\rm ^4He}=2\sigma^{\rm
^4He}/(A\sigma^{\rm D})=F_2^{\rm ^4He}/F_2^{\rm D}$ are presented
at Fig.\ref{fig1} a). The nuclear momentum distribution is taken
from~\cite{ciofisimula}, the nucleon SF $F_2$ is taken
from~\cite{phlett04}. The first term ($R^{\rm ^4He}(IA)$), which
results from the Fermi motion in usual $3$D-space, is presented by
the monotone increasing dashed curve. The second term ($\Delta
R^{\rm ^4He}$) resulted from the Fermi motion along the time axis
is depicted by the point-dashed curve. Since $M_A<E_{\rm
N}-E_{A-1}$ and $F_2^{\rm N}(x_{\rm N})/dx_{\rm N}<0$ it gives
negative contribution to $F_2^A$. This term puts the ratio $R^{\rm
^4He}$ bellow unity in the middle $x$ region and, therefore,
produces the EMC-effect. The full curve presents sum of these
contributions. It clearly falls to the experimental data within
the experimental errors with rather good accuracy (see
Fig.\ref{fig1} b)).

\begin{figure}
\epsfxsize=70mm \epsfbox{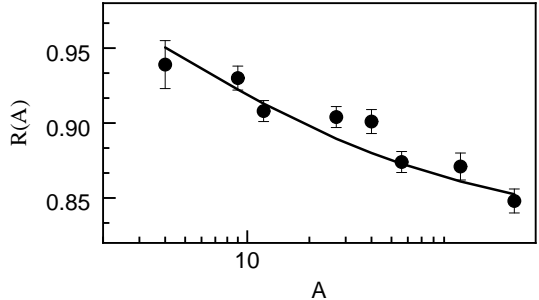}
\caption{\label{fig2} A-dependence of the nuclear to deuteron structure functions ratio 
calculated at $x=0.6$.
The full line is the calculation according to the Eqs.~(\ref{Aapp}) and (\ref{Ta}).
The experimental values are shown by the full circles~\protect\cite{gomez}.}
\end{figure}

Thus, we have obtained that the $\tau$-shift affects on the
observable quantities in DIS. It broadens the bound nucleon
localization area in $4$D-space, what leads to the observation of
the EMC-effect. This phenomenon resembles the one predicted by the
hypothesis of the $x$-rescaling and $Q^2$-rescaling. However,
these models assume real change of the bound nucleon structure,
while in the present picture the EMC-effect results from the
measurement uncertainty due to the the $4$D-Fermi motion. This is
the key difference between these models and the presented here
picture of the EMC-effect.
The binding approach that use separation energy as rescailing
parameter clearly underestimates the EMC-effect at the large $x$ region.
The Eq.(\ref{F2A}) shows that amplitude of the EMC effect  is defined by the
geometrical properties of nuclei, which expressed by the
parameter $\Delta^{N}_{A}$ in the momentum space. It is worth noting that $A$-dependence 
of this quantity explains the $A$-dependence of the EMC-effect 
observed on the experiment~\cite{gomez}. It can be qualitatively shown 
with the help of an approximated form of $\Delta^{N}_{A}$. Assuming weak $A$-dependence 
of the binding energy ($\epsilon_A\simeq \epsilon_{A-1}$) and small relative 
3-momentum of the bound nucleons ($\langle {\bf p}^2 \rangle \ll m^2$), 
we can rewrite mean value of $\Delta^{N}_{A}$ in the form:
\begin{equation}
\langle \Delta^{N}_{A}\rangle=\epsilon_A-\langle T \rangle_{A} \frac{A}{A-1}
\label{Ta}\end{equation}
where $\epsilon_{A}\simeq (M_A-A m)/A$ is the binding energy of the nucleon and 
$\langle T \rangle_{A}\langle {\bf p}^2 \rangle/(2 m)$ is the mean kinetic energy 
of the bound nucleon. 
Thus, the amplitude of the deviations from unity of the structure functions ratio 
is defined by the binding enegry $\epsilon_A$ and mean kinetic
energy $\langle T \rangle_A$ of the nucleon bound in
the nucleus $A$. The binding energy weakly depends on $A$, since then 
$A$-dependence of the EMC-effect is defined mainly by the nucleon kinetic energy. 
Due to the uncertainty relation $\langle T \rangle_A$ 
is proportional to the mean nuclear density  
$\langle T \rangle_A \propto \rho_A^{2/3}$, what explains the observation 
made in SLAC experiment~\cite{gomez}.
Taking into account that the mean density in the central part of the nucleus $\rho_c$ and on 
the surface $\rho_c$ weakly depend on $A$ and related as   
$\rho_s/\rho_c\simeq 0.02$, we can derive the eplicite expression for $A$-dependence 
of the nucleon mean kinetic energy:
\begin{equation}
\langle T \rangle_A \simeq \langle T_c \rangle \left(1-0.98 \frac{N_s(A)}{A}\right).
\label{Aapp}\end{equation}
Here $N_s(A)$ is the number of nucleons on the surface of the nucleus, $\langle T_c \rangle$ 
is the mean kinetic energy of the bound nucleon in the central part of the nucleus 
($\langle T_c \rangle \simeq $).
This expression corresponds to the world experimental data fit for nuclear DIS made
in the paper~\cite{sm95}.
Numerical calculation of this estimation is in a good agreement with experimental data (see Fig.\ref{fig2}). 

In conclusion, the $4$D-Fermi motion plays two-fold role in the nuclear DIS. From
the one hand the uncertainty in measurement does not allow one to
exactly detect nucleon structure in nuclear experiments, so that
the information about nucleon structure extracted from the nuclear
data strongly depends on the model used for description of the
nucleon. It leads to the well known difficulties in extraction of
the neutron structure functions from the nuclear
experiments~\cite{phlett04,ThomasF2n}. From the other hand,
Eq.~(\ref{F2A}) shows that the nucleon structure function
derivative defines the pattern of the nuclear to deuteron DIS
cross-sections ratio. Thus, the $4$D-Fermi motion enables direct
access to the information about nucleon structure dynamics. It is
worth noting that the proton SF has a large negative slope
 at small $x$, what can lead to the deviation of $R^A$ from unity
 producing in that way the effect that coincides with the nuclear
 shadowing.
Thus, the $4$D-Fermi motion can give an universal explanation of
the EMC-effect, anti-shadowing, and shadowing, what can be
confirmed by further studies at small $x$ region.

An another important outcome is the direct link between
sub-nucleon structure dynamics and space-time geometry at
microscopic distances. Within the
covariant $4$D approach the space-time is considered as the metric
$4$-space, the $4$D Fermi-motion in which leads to the EMC-effect.
The covariant 3D approaches consider 4D
space-time as a $3+1$ foliated manifold with the time as a
factor-space and the metric $3$-space as a layer. The $3$D Fermi-motion 
is shift of the particle position in the layers with move along the factor space.
To take into account the effects of the Fermi-motion along the time axis,
one has to introduce dynamical distortions particle move. Thus, 
within a such approach the EMC-effect can be explained by dynamical change of
bound nucleon structure. For example, the central mechanism in
QMC~\cite{thomas03} and chiral soliton models~\cite{miller03} to
explain the EMC effect is the scalar attractive interaction, which
modifies the quark distribution in the nucleon. 
Influence of the Fermi-motion along time-axis on the observable
quantities has scalar character. In that way, the scalar field in
the covariant $3$D approach can be considered as a
three-dimensional dynamical realization of the time component of
the $4$D-Fermi motion.

In summary, the explicitly covariant $4$D approach provides a
consistent description of the nuclear short-range structure in
terms of the nucleon structure and four-dimensional Fermi motion
of the nucleons. The Fermi motion along the time axis broadens the
bound nucleon localization area, what lead to the observation of
the nucleon structure change in nuclei -- EMC effect.
The amplitude of the EMC-effect is defined by the binding and kinetic 
energy of the bound nucleons, thus the $A$-dependence of the EMC-effect 
can be explained by decrease with $A$ of the number of surface nucleons 
that have much smaller kinetic energy than nucleons bound in the 
central part of the nucleus.   
The pattern of the
effect is defined by dynamics of the nucleon structure. Thus,
the obtained results are in qualitative consistence with the
conclusion of the Drell-Yan experiments~\cite{DY}. Detailed
quantitative comparison need further study with extension of the
presented formalism to the Drell-Yan process.

I would like to thank  U.~Mosel, G.I.~Smirnov, and
H.~Toki for useful discussions.

\end{document}